# Interplanetary and Geomagnetic Consequences of Interacting CMEs of 13 – 14 June 2012

**Nandita Srivastava**[1] · **Wageesh Mishra**[2] · **D. Chakrabarty**[3]

**Abstract** We report on the kinematics of two interacting CMEs observed on 13 and 14 June 2012. Both CMEs originated from the same active region NOAA 11504. After their launches which were separated by several hours, they were observed to interact at a distance of 100 $R_{\odot}$ from the Sun. The interaction led to a moderate geomagnetic storm at the Earth with $D_{st}$ index of approximately, -86 nT. The kinematics of the two CMEs is estimated using data from the *Sun Earth Connection Coronal and Heliospheric Investigation* (SECCHI) onboard the *Solar Terrestrial Relations Observatory* (STEREO). Assuming a head-on collision scenario, we find that the collision is inelastic in nature. Further, the signatures of their interaction are examined using the *in situ* observations obtained by *Wind* and the *Advance Composition Explorer* (ACE) spacecraft. It is also found that this interaction event led to the strongest sudden storm commencement (SSC) ($\approx$ 150 nT) of the present Solar Cycle 24. The SSC was of long duration, approximately 20 hours. The role of interacting CMEs in enhancing the geoeffectiveness is examined.

**Keywords:** Coronal mass ejections, STEREO, Heliospheric Imagers

[1] Udaipur Solar Observatory, Physical Research Laboratory,
Udaipur, India -313001
email: nandita@prl.res.in

[2] CAS Key Laboratory of Geospace Environment,
Department of Geophysics and Planetary Sciences,
University of Science and Technology of China, Hefei, China
email: wageesh@ustc.edu.cn

[3] Space and Atmospheric Sciences Division, Physical
Research Laboratory, Ahmedabad-380009, India
email: dipu@prl.res.in

## 1. Introduction

After the launch of twin *Solar TErrestrial Relations Observatory* spacecraft (STEREO: Kaiser *et al.* 2008), the data from the *Sun Earth Connection Coronal and Heliospheric Investigation* (SECCHI: Howard *et al.* 2008) have enabled us to continuously image CMEs from their lift-off in the corona up to the Earth and beyond (Davies *et al.*, 2009; Harrison *et al.*, 2012; Liu *et al.*, 2013; Möstl *et al.*, 2015; Vemareddy and Mishra, 2015). These observations have also revealed direct evidence of CME–CME interaction when they are launched in close succession in the same direction (Harrison *et al.*, 2012; Lugaz *et al.*, 2012; Shen *et al.*, 2012; Mishra, Srivastava, and Chakrabarty, 2015). In fact, CME interactions are now commonly observed, in particular, around solar maximum when the occurrence of CMEs is larger in number. A number of studies pertaining to understanding of the individual cases of interacting CMEs have been reported highlighting the nature of the interaction and/or collision and their signatures for example by Harrison *et al.* 2012, Liu *et al.* 2012, Möstl *et al.* 2012, Temmer *et al.* 2012, Lugaz *et al.* 2012. Based on these observations of interacting CMEs, some of the crucial questions have been aptly addressed as to whether the nature of interaction is elastic, inelastic, or super-elastic? (Lugaz *et al.*, 2012; Shen *et al.*, 2012; Mishra and Srivastava, 2014; Mishra, Srivastava, and Chakrabarty, 2015; Mishra, Wang, and Srivastava, 2016). Further, using *in situ* observations, it may also be possible to answer the question, under what conditions, interacting CMEs lead to a merged or a separate structure. One of the important questions is whether the interaction of CMEs leads to enhanced geoeffectiveness as indicated by Farrugia *et al.* (2006). If so, what are the distinct signatures of the same? Interaction of CMEs also has bearing on the prediction of space weather as the kinematics of CMEs change after interaction. For this purpose one also needs to understand the pre- and post-interaction kinematics which can influence the resulting geoeffectiveness.

In this study, we present the evolution, propagation and interaction of two CMEs launched on 13 and 14 June 2012 as they travelled in the inner heliosphere and reached the Earth. The observations reveal that the launch times of the CMEs were separated by about 24 hr. The observations also reveal that the CMEs of 13 and 14 June were directed towards the Earth and their initial speed values indicate their probable interaction as they propagated out in the heliosphere. In this regard, the event provides an excellent opportunity for us to understand the interaction process in details. To achieve this objective, we have estimated the 3D kinematics of the two CMEs using STEREO/SECCHI observations, and the distance from the Sun at which they interacted. We have also calculated the true mass of the interacting CMEs and their momentum to reveal the type of collision, and estimated momentum and energy transfer during the collision phase. Taking into account the propagation characteristics, the type of collision and energy transfer properties of the two CMEs during the interaction were estimated. We also examined the *in situ* data of the tracked CME features. The geomagnetic consequence of the interacting CMEs are very distinct and have been studied in detail. The results and conclusions are presented in the final section.

## 2. Observations

For the CME–CME interaction event of 13 – 14 June 2012, we analyzed data from the SECCHI suite (Howard *et al.*, 2008) onboard NASA's twin STEREO (A and B) mission. The SECCHI package includes the *Extreme Ultraviolet Imager* (EUVI), two coronagraphs (COR1 and COR2), and two Heliospheric Imagers (HI1 and HI2)(Eyles *et al.* 2009). These instruments can track a CME from near the Sun to the Earth and further into the heliosphere. The field of view (FOV) of the EUVI is 1–1.7 $R_{\odot}$, of COR1 is 1.5–4 $R_{\odot}$ and of COR2, is 2.5–15 $R_{\odot}$. The FOVs of the EUVI and CORs are centered on the Sun. On the other hand, HI1 and HI2 have their optical axis aligned in the ecliptic plane and are off-centered from the Sun at solar elongation of 14° and 53.7°, respectively. The field of view of HI1 and HI2 is 20° and 70°, respectively. Thus using SECCHI/STEREO instrument data, a CME can be tracked from 0.4 to 88.7° elongation. For quick examination, we also used the coronagraph observations by the Large Angle and Spectrometric Coronagraphs (LASCO: Brueckner et al., 1995) onboard the Solar and Heliospheric Observatory (SOHO). During 13 – 14 June 2012, STEREO A and B were 117° westward and 118° eastward from the Sun–Earth line at a distance of 0.96 AU and 1.0 AU from the Sun, respectively. The *in situ* properties of interacting CMEs were studied using the OMNI database which includes data recorded by instruments on board *Wind* (Lepping *et al.*, 1995; Ogilvie *et al.*, 1995) and *Advance and Composition Explorer* (ACE) spacecraft (Stone *et al.*, 1998).

### 2.1. Analysis of CMEs of 13–14 June 2012

A partial halo CME (CME1) at 13:25 UT with a projected speed of 630 km $s^{-1}$ was recorded by LASCO coronagraphs on 13 June 2012. On the next day, *i.e.* 14 June, another halo CME (CME2) was recorded around 24.8 hr after the launch of CME1, having a projected speed of 990 km $s^{-1}$ at 14:12 UT. The two CMEs are shown in Figure 1. As CME1 had a slower speed than CME2 and both seemed to propagate in the earthward direction, the observations suggest the probability of their interaction in the heliosphere. Both CMEs originated from the same active region, *i.e.* from NOAA AR 11504. CME1 was associated with an M1.2 flare which occurred at around S16 E18 location on 13 June 2012 and CME2 was associated with an M1.0 flare in the same active region and occurred on 14 June. The separation angle of the STEREO spacecraft during 13–14 June 2012 was large, *i.e.* 126°, therefore, in order to estimate the 3D kinematics of the CMEs, the Graduated Cylindrical Shell (GCS model: Thernisien, Howard, and Vourlidas, 2006) was used for 3D reconstruction of the CMEs. This model represents the large scale flux rope structure of CMEs. It consists of a tubular section which forms the main body of the structure attached to two cones that form the legs of the CME. The resulting shape resembles that of a hollow croissant. For the present case, we have applied the GCS forward fitting model to the contemporaneous images of the CMEs obtained from the SECCHI/COR2-B, SOHO/LASCO-C3, and SECCHI/COR2-A coronagraphs as shown in the images overlaid with the fitted GCS wireframed contour (Figure 2). From the fitting of the GCS model, we estimate the propagation direction of CME1 along E15 S30 at 10.9 $R_{\odot}$. The propagation direction for the following CME2 was along E02 S25 at 13.5 $R_{\odot}$. In addition to the propagation direction, the best visual GCS fitting gives a half angle of 22.5°, a tilt angle of 70°, and an aspect ratio of 0.55 for CME1. The half angle, tilt angle and aspect ratio for CME2 is 30°, 70°, 0.6, respectively. At around 11 $R_{\odot}$, the 3D speed of CME1 is estimated as 560 km s$^{-1}$ and for CME2 as 900 km s$^{-1}$. The directions and speeds of the CMEs suggest that they possibly collide during their heliospheric evolution. Using SECCHI/HI observations, we determined the distance from the Sun at which the interaction took place and also the nature of collision. We also attempted to identify distinct CME structures in the *in situ* data taken at L1 point and used them to estimate the arrival time of the interacting CMEs.

#### *2.1.1. 3D Reconstruction of Interacting CMEs in HI fov:*

The CMEs of 13–14 June 2012 were well observed also in the HI-A and HI-B field of views of STEREO also. For the tracking of CME features, a minimum background image was created from a sequence of HI images. We constructed the time-elongation maps, conventionally called J-maps, (Sheeley *et al.*, 2008; Davies *et al.*, 2009) using the running difference images of HI-1 and HI-2. The details of the procedure to construct the J-maps and to derive the distance from the measured elongation angles have been described in Mishra, Srivastava, and Chakrabarty (2015); Mishra, Wang, and Srivastava (2016). These J-maps reveal the kinematic evolution of these CMEs and are shown in Figure 3. The positively inclined bright features in the J-maps correspond to the enhanced density structure of the CMEs. The J-maps for 13–14 June events indicate that enhanced brightness features came in close contact with each other and appear to merge around 25° elongation. Both features were tracked further up to 35° elongation.

On the basis of earlier studies regarding the relative performance of the reconstruction methods (Lugaz, 2010; Liu *et al.*, 2013; Mishra, Srivastava, and Davies, 2014), we applied the stereoscopic self-similar expansion (SSSE) (Davies *et al.*, 2013) method to the J-maps of the interacting CMEs in the present study. The SSSE method is expected to yield better results as the CMEs were observed in HI field of view of both STEREO-A and B. To implement the SSSE method, we require an appropriate value of the cross-sectional angular half width ($\lambda$) of the CMEs as an input. Earlier studies have also revealed that use of different values of $\lambda$ with the SSSE method give different estimates of the kinematics of CMEs (Liu *et al.*, 2013; Mishra, Srivastava, and Davies, 2014). It has been observed that for CMEs that are Earth-directed when STEREO spacecraft are behind the Sun, the SSSE method should be implemented with a value of $\lambda$ as 90° (Liu *et al.*, 2013, 2014; Mishra, Srivastava, and Singh, 2015; Vemareddy and Mishra, 2015). In our case, the CMEs are Earth-directed and therefore the SSSE method is implemented with a value of $\lambda$ as suggested in earlier studies. The estimated kinematics by implementing the SSSE method on the derived time-elongation profiles of these CMEs is shown in Figure 4 and has been used to understand the collision phase of CMEs. As described in an earlier article (Mishra, Srivastava, and Chakrabarty 2015), we define the collision phase as the interval during which the two CMEs come in close contact with each other and show opposite trends of acceleration relative to one another until they attain an approximately equal speed or their trend of acceleration is reversed. The collision of the two interacting CMEs occurred between 8:40 UT to 15:50 UT in a 7.2 hr span on 15 June 2012. At the beginning of the collision, the tracked feature of CME1 was at around 105 $R_{\odot}$ and that of CME2 at around 100 $R_{\odot}$. During the collision, they traveled a distance of around 25 $R_{\odot}$ before reaching an approximately equal speed. The collision led to an acceleration of the preceding CME1 from 590 km $s^{-1}$ to 680 km $s^{-1}$ and a deceleration of the following CME2 from 865 km $s^{-1}$ to 680 km $s^{-1}$.

## 2.2. Momentum, Energy Exchange and Nature of the Collision

To estimate the momentum exchange during the collision of the CMEs, it is required to estimate the true masses of the two CMEs. Assuming that the CME observed in white-light is due to Thomson scattered photospheric light from the electrons in the CME (Minnaert, 1930; Billings, 1966; Howard and Tappin, 2009), the recorded scattered intensity can then be converted into the number of electrons, and hence the mass of a CME can be estimated, assuming a completely ionized corona with a composition of 90% hydrogen and 10% helium. In earlier studies (Munro *et al.*, 1979; Poland *et al.*, 1981; Vourlidas *et al.*, 2000), the mass of a CME was calculated assuming the CME location in the observer's plane of sky. We use the method of Colaninno and Vourlidas (2009) which is based on the Thomson scattering theory, to estimate the true propagation direction and true mass of both CMEs. For this purpose we used the simultaneous image pair of SECCHI/COR2 and estimated the masses of CME1 and CME2 to be $8.4 \times 10^{12}$ kg and $9.2 \times 10^{12}$ kg, respectively. The mass ratio of the interacting CMEs is approximately 1.1.

Although, we have estimated the true mass, this also involves uncertainties. A straightforward uncertainty arises due to the assumption that the mass of a CME is concentrated on the plane-of-sky. However, a CME is a three dimensional structure with a significant depth along the line-of-sight. It has been reported earlier that such an assumption leads to underestimation of the CME mass by 15% (Vourlidas *et al.*, 2000). We made several independent mass measurements for this event and found that the values are within 20% of the measured ones. Further, the role of the uncertainty in the mass estimation has been studied to understand the variation in the coefficient of restitution which has been found to be negligible in deciding the nature of the collision (Shen *et al.*, 2012; Mishra, Wang, and Srivastava, 2016). This is expected as our approach constrains the momentum conservation while modifying the observed post-collision speeds (Mishra and Srivastava, 2014). Therefore, in this study we did not assess the effect of uncertainties in the mass. Significant momentum exchange takes place during the interaction, with an increase in the momentum of the preceding and decrease in the momentum of the following CME. In the present case, the momentum of CME1 increased by 57% and that of CME2 decreased by 24% after the collision.

For our analysis we consider that after crossing the FOV of the COR2 coronagraph, during the collision of CME1 and CME2, their estimated true masses ($M_1$ and $M_2$) remained constant. Using the estimated kinematics before and after the collision (Figure 4) and their true masses, the coefficient of restitution has been calculated. The observed velocity of CME1 and CME2 before the collision is estimated as $(u_1, u_2) = (590,\ 865)$ km s$^{-1}$ and the observed velocity of CME1 and CME2 after the collision is $(v_1, v_2) = (680,\ 680)$ km s$^{-1}$. To understand the nature of the collision of the two CMEs, we estimate the coefficient of restitution ($e$) of the colliding CME1 and CME2 following the formulations described in Mishra and Srivastava (2014); Mishra, Srivastava, and Singh (2015). The coefficient of restitution measures the bounciness of the collision and is defined as the ratio of their relative velocity of separation to their relative velocity of approach. In this particular case, the coefficient of restitution ($e$) is estimated to be 0.0, *i.e.* the collision is perfectly inelastic in nature. Consequently the kinetic energy of the system after the collision is found to be lower than that before the collision.

## 3. *In situ* Properties and Time of Arrival of the Interacting CMEs of 13 – 14 June 2012

We analyzed the *in situ* data associated with the 13-14 June, interacting CMEs using the OMNI database to identify interplanetary signatures of different features of the CMEs. Figure 5 shows the magnetic field and plasma measurements during 00:00 UT on 16 June to 00:00 UT on 18 June. The arrival of a forward shock (labeled S1) marked by a sudden enhancement in speed, temperature, and density is observed at 08:42 UT on June 16. This is followed by an ICME structure (ICME1) for approximately 12 hrs. The second shock S2 is marked by a sharp and huge increase in density indicating a pile-up or compression of the plasma as the shock passes through the cloud. S2 was observed at 21:40 UT on 16 June. Based on the signatures of ICMEs that are expected to be observed in *in situ* observations (Zurbuchen and Richardson, 2006), the region bounded between approximately 21:40 UT on 16 June and 21:00 UT on 17 June is identified as the second ICME structure (ICME2).

During the passage of ICME1, the magnetic field was quite strong ($\approx 10$ nT), while plasma $\beta < 1$ was largely lower than unity. In the case of ICME2, the magnetic field is even stronger (40 nT). A gradual decrease in magnetic field, density and velocity is observed until 21:00 UT on 17 June. During the passage of this CME, plasma $\beta < 1$, temperature is also low ($\approx 10^4$ K) and the density high. These signatures are suggestive of the passage of a magnetic cloud as defined by Burlaga et al. (1981). It is further observed that the temperature does not show any distinct variation during the interaction of the two CMEs contrary to our previous studies (Mishra and Srivastava, 2014; Mishra, Srivastava, and Chakrabarty, 2015) where the interaction of the CMEs was represented by a high temperature region associated with the first CME, which is higher than usually found for a normal isolated CME (Zurbuchen and Richardson, 2006).

Thus, examining the *in situ* observations, we can mark the arrival of two distinctly different regions as far as the CME–CME interaction is concerned in the present case. The first part appears to be the arrival of the first CME marked by the increase of the plasma parameters at 8:42 UT on 16 June but steady for some time. Then the second CME at 21:40 UT on 16 June probably pierces through the first one creating a small region where the density falls off very rapidly. The two peaks in SYM-H (Iyemori, 1990) at 22:40 UT on 16 June and at 01:05 UT on 17 June interspersed by a dip probably correspond to the arrival of the faster CME in which a rarefied region is embedded. The estimated speeds of two shocks marked as S1 and S2 were 450 and 485 km $s^{-1}$, respectively. The proton temperature ratio $Tp_{down}/Tp_{up}$ was approximately 0.04 for the second shock as compared to 1.26 for the first shock.

## 4. Geomagnetic Consequences of the Interacting CMEs of 13 − 14 June 2012

The CMEs on 13 − 14 June 2012 were quite normal CMEs in terms of their initial speeds. They resulted in a single moderate ($D_{st} \approx$ -86 nT) and long lasting geomagnetic storm. However, the magnitude of the sudden storm commencement (SSC) was exceptionally high ($\approx$ 150 nT) at 21:40 UT on 16 June. This calls our attention to study the impact of the interaction of the two CMEs on the terrestrial magnetosphere-ionosphere system in detail.

### 4.1. Sudden Storm Commencement

As mentioned above, the resulting geomagnetic storm is important because of its intense magnitude, particularly of the associated SSC. Generally the time of the SSC denotes the arrival of the interplanetary shock and its strength. Previous studies based on a statistical analysis have shown that the occurrence rate of SSCs is less than 5% for amplitudes larger than 50 nT and less than 1% for amplitudes larger than 100 nT (Araki, 2014). Further, generally large amplitude SSCs tend to occur during the declining phase of the solar activity, however, the present case is an exception as it occurred during the ascending phase of Solar Cycle 24. The SYM-H index rose from 39 nT to 150 nT, during 16 June 2012, 20:20 UT to 20:47 UT, the rise time being 27 minutes. This puts the observed SSC as the strongest observed in Solar Cycle 24 and also as one of the most intense events if one considers the past events examined by Araki et al. (2014). Further, SSCs with amplitude larger than 100 nT are extremely rare (less than 1%) and the rise time is usually 3 to 4 min (Maeda *et al.*, 1962). In this regard, the amplitude of this SSC is very large and the rise time is unusually long. The present case due to its large variation with time, strongly suggests the strengthening of the shock which occurred due to the interaction of the two successive CMEs. This is unique for three reasons. First, is its long duration (more than 20 hr); second, its high peak magnitude ($\approx$ 150 nT); and third, it occurred in three distinct steps with a small rise to start with and two peaks with a sharp fall in between. It is also interesting to point out that during this time the y-component of interplanetary electric field (IEFy) was negative, which means that the z-component of the interplanetary magnetic field (IMF), $B_z$, was northward during this interval, thereby implying that its time of occurrence is prior to the main phase of the storm and hence confirms that the feature is an SSC.

Also, in an attempt to understand the role of interacting CMEs for the unique SSC event, we observed that solar wind density increased three times from 20 to 60 cm$^{-3}$ and the velocity from 300 to 550 km s$^{-1}$ during this interval, which corresponds to distinct steps manifested in SYM-H values. This further suggests that solar wind ram pressure related shocks have contributed significantly to the SSC.

The IP shock properties as derived from Wind data and catalogued at http://ipshocks.fi/ reveal that for the second shock (S2), the value of $\theta$, *i.e.* the angle between the shock normal and the magnetic field lines upstream is 60° as compared to 15°, for the first shock (S1). Here, the values of $\theta$ indicate that shock S1 is quasi-parallel ($\theta < 45°$) and therefore should be associated with extended foreshock regions, whereas the shock transitions are typically more gradual; the jumps in the solar wind plasma parameters and in the magnetic field magnitude are less significant than at quasi-perpendicular shocks (Burgess *et al.*, 2005; Kruparova *et al.*, 2013), where $\theta > 45°$.

### 4.2. Main Phase of Geomagnetic Storm

As may be noted from Figure 5, the main phase of the geomagnetic storm lasted for more than 16 hr and the recovery phase was also quite long (approximately 72 hr). *In situ* observations indicate the arrival of two shocks and a merged structure for the two events. In particular, the OMNI *in situ* data reveal a weak shock at 09:03 UT with a small ICME followed by a stronger shock at 19:34 UT and a prolonged ICME. It is also noted that the intensity of the geomagnetic storm, $D_{st}$ reached a peak value of ≈ -86 nT at 14:00 UT and maintained the moderate level of −50 nT for 14 hr. The event suggests a resemblance with that described by Lugaz and Farrugia (2014), wherein they reported a long-duration isolated event that resulted from the merging of two CMEs with peak $D_{st}$ reaching ≈ 150 nT and remaining at moderate values (below 50 nT) for 55 hr. It may also be noted that we do not observe any signature of a distinct interaction region in *in-situ* data for 13 − 14 June event, as was reported in the case of 9–10 November 2012 interacting CMEs (Mishra, Srivastava, and Chakrabarty, 2015).

## 5. Summary and Conclusion:

Although the interacting CMEs of 13 − 14 June 2012 appear to be quite normal in terms of speeds and associated flares and the resulting geomagnetic storm is also moderate with $D_{st}$ attaining ≈ -86 nT value, the interaction event is quite unique in terms of its geomagnetic consequence. The two CMEs interacted at a distance of 100 $R_\odot$ from the Sun and reached the Earth as a merged structure. The arrival of the CMEs is marked by an enhanced SSC with a peak of 150 nT. The magnitude of this SSC is the highest recorded in Solar Cycle 24. The duration of the rise time of this SSC is also unusually high, of the order of several hours due to the strengthening of the shock owing to the interaction of the CMEs. The merged structure led to a single step moderate storm whose duration was unusually long, both for the main phase ($\approx 16$ hr) and the recovery phase ($\approx 72$ hr). Contrary to the present knowledge that the strong SSCs occur during the descending phase of the solar cycle, the CMEs of 13 – 14 June 2012 are remarkable as their interaction led to the strongest SSC in the ascending phase of Solar Cycle 24.

**Acknowledgments** We acknowledge the UK Solar System Data Centre for providing the processed level-2 STEREO/HI data. The *in situ* measurements of solar wind data from ACE and Wind spacecraft were obtained from NASA CDAweb (http://cdaweb.gsfc.nasa.gov/). W. M. is supported by the Chinese Academy of Sciences (CAS) President's International Fellowship Initiative (PIFI) grant No. 2015PE015.

**Disclosure of Potential Conflicts of Interest:** The authors declare that they have no conflicts of interest.

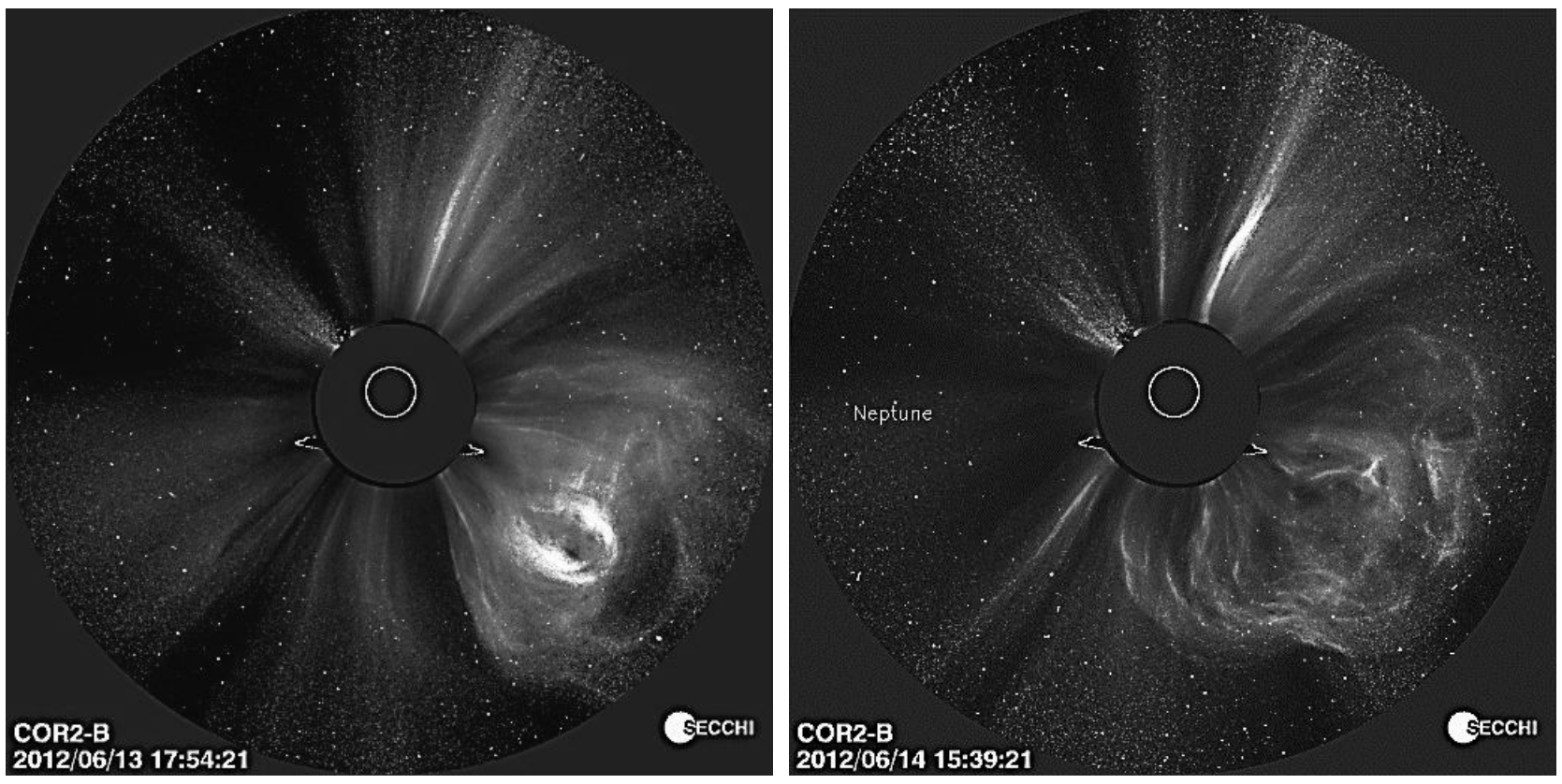

**Figure 1.** The two interacting coronal mass ejections of 13 and 14 June 2012 observed as partial halos by COR2-B coronagraph in the left and right panels, respectively.

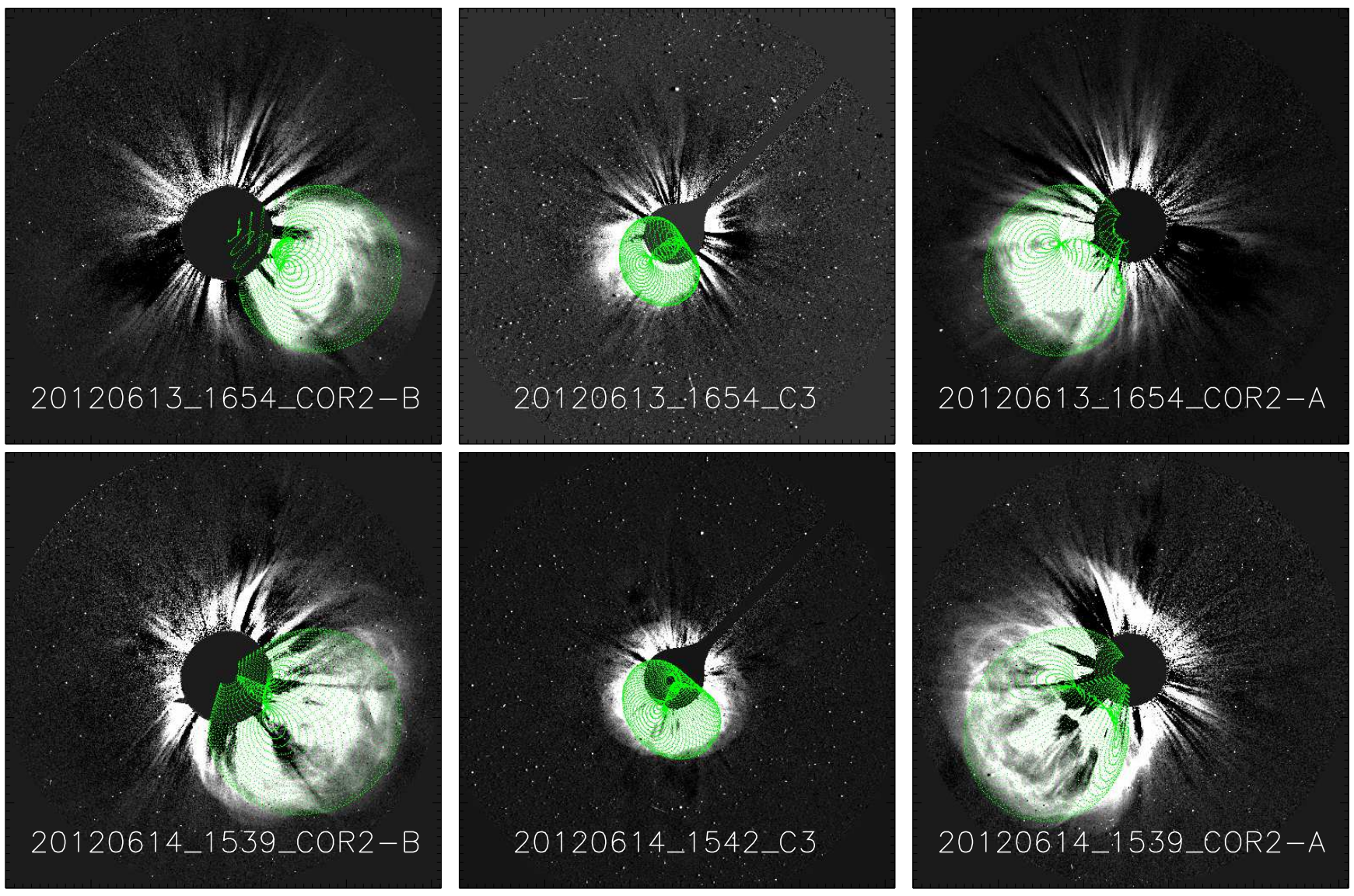

**Figure 2.** The GCS wired model overlaid on the contemporaneous images observed for the two CMEs, upper panel (CME1) around 16:54 UT on 13 June and lower panel (CME2) around 15:39 UT on 14 June. For both CMEs the fitting of the model is done for the three images recorded by COR2-B (left), LASCO-C3 (middle), COR2-A (right) images.

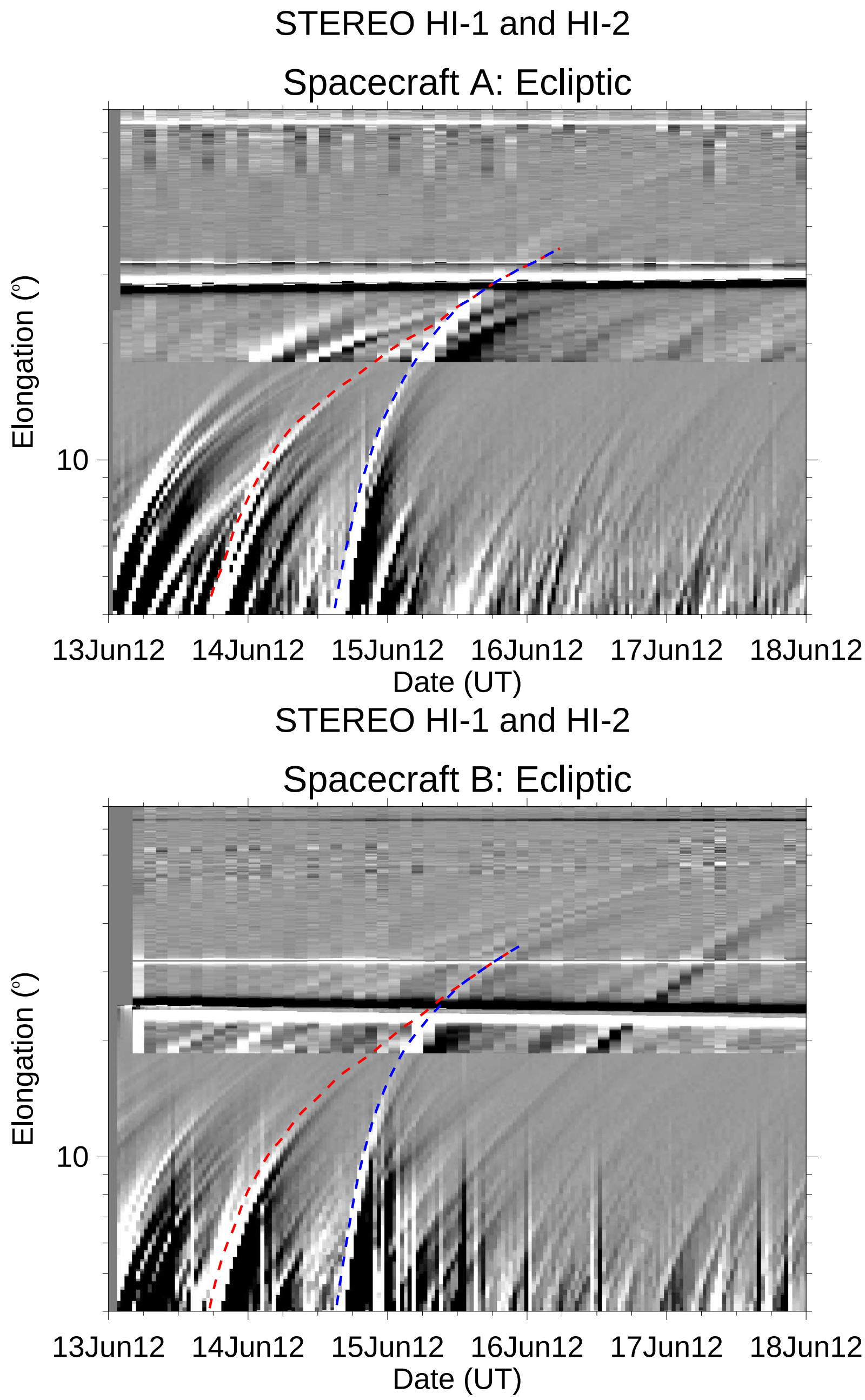

**Figure 3.** Time-elongation maps (J-maps) using the COR2 and HI observations of STEREO/SECCHI spacecraft during interval of 13 – 14 June 2012 is shown. The features corresponding to leading edges of CME1 and CME2 are tracked and overplotted on the J-maps.

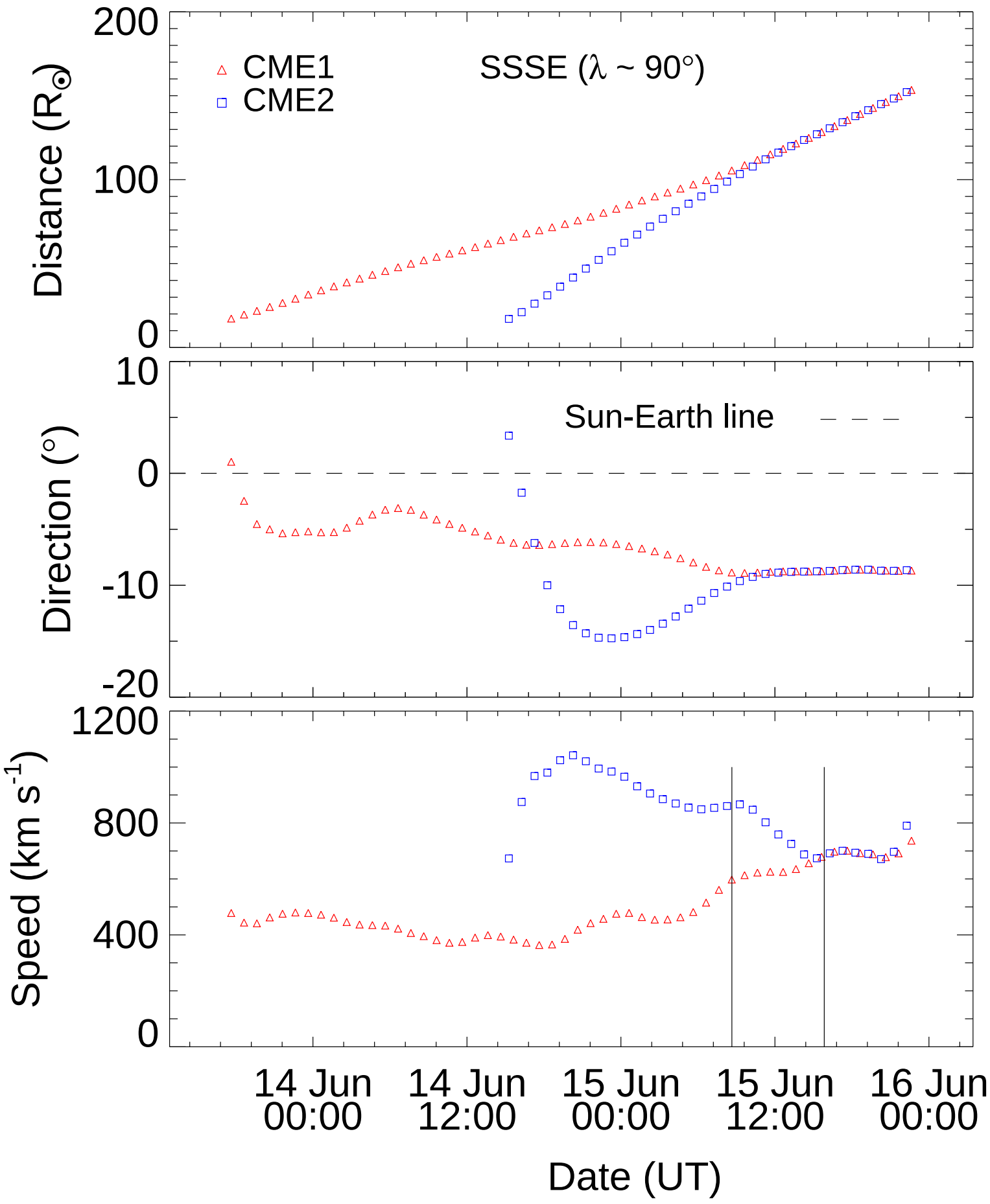

**Figure 4.** The plots show the distance and speed estimated for different tracked features of the two CMEs.

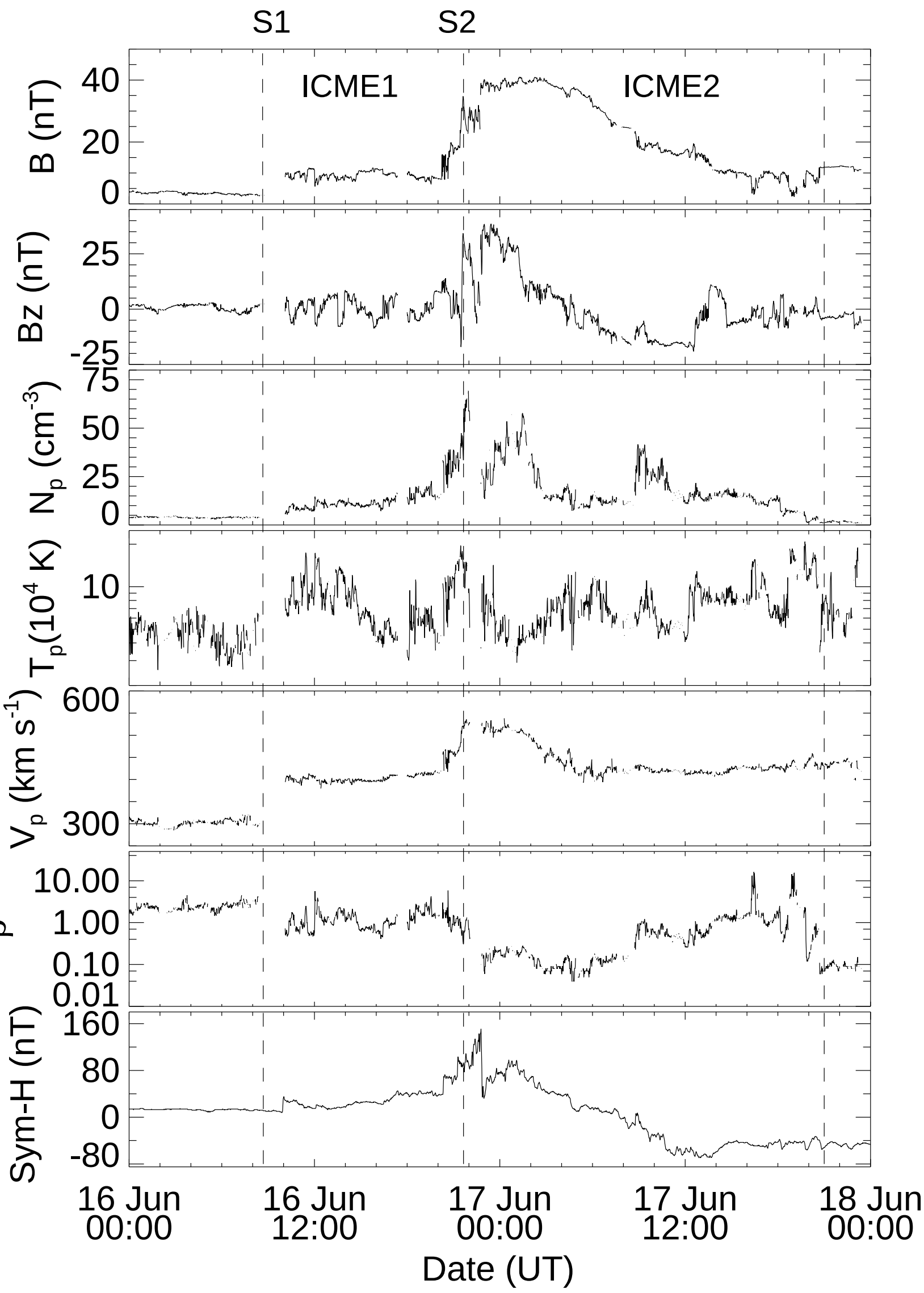

**Figure 5.** From top to bottom, total magnetic field magnitude, z-component of magnetic field, proton density, proton temperature, proton speed, plasma beta ($\beta$), and SYM-H, is shown for the time interval of 00:00 UT on 16 June to 00:00 UT on 18 June. From the left, the first, second, and third vertical (dashed) lines mark the arrival of shock (S1) associated with CME1, shock (S2) associated with CME2, and the trailing boundary of ICME2, respectively.

## References

Araki, T.: 2014, Historically largest geomagnetic sudden commencement (SC) since 1868. *Earth Planets Space* **66**, 164. DOI. ADS.

Billings, D.E.: 1966, *A guide to the solar corona*, Academic Press, New York, 150. ADS.

Brueckner, G.E., Howard, R.A., Koomen, M.J., Korendyke, C.M., Michels, D.J., Moses, J.D., Socker, D.G., Dere, K.P., Lamy, P.L., Llebaria, A., Bout, M.V., Schwenn, R., Simnett, G.M., Bedford, D.K., Eyles, C.J.: 1995, The Large Angle Spectroscopic Coronagraph (LASCO). *Solar Phys.* **162**, 357. DOI. ADS.

Burgess, D., Lucek, E.A., Scholer, M., Bale, S.D., Balikhin, M.A., Balogh, A., Horbury, T.S., Krasnoselskikh, V.V., Kucharek, H., Lembège, B., Möbius, E., Schwartz, S.J., Thomsen, M.F., Walker, S.N.: 2005, Quasi-parallel Shock Structure and Processes. *Space Sci. Rev.* **118**, 205. DOI. ADS.

Burlaga, L., Sittler, E., Mariani, F., & Schwenn, R. 1981, *J. Geophys. Res.*, 86, 6673

Colaninno, R.C., Vourlidas, A.: 2009, First Determination of the True Mass of Coronal Mass Ejections: A Novel Approach to Using the Two STEREO Viewpoints. *Astrophys. J.* **698**, 852. DOI. ADS.

Davies, J.A., Harrison, R.A., Rouillard, A.P., Sheeley, N.R., Perry, C.H., Bewsher, D., Davis, C.J., Eyles, C.J., Crothers, S.R., Brown, D.S.: 2009, A synoptic view of solar transient evolution in the inner heliosphere using the Heliospheric Imagers on STEREO. *Geophys. Res. Lett.* **36**, 2102. DOI. ADS.

Davies, J.A., Perry, C.H., Trines, R.M.G.M., Harrison, R.A., Lugaz, N., Möstl, C., Liu, Y.D., Steed, K.: 2013, Establishing a stereoscopic technique for determining the kinematic properties of solar wind transients based on a generalised self-similarly expanding circular geometry. *Astrophys. J.* **776**, 1.

Eyles, C.J., Harrison, R.A., Davis, C.J., Waltham, N.R., Shaughnessy, B.M., Mapson-Menard, H.C.A., Bewsher, D., Crothers, S.R., Davies, J.A., Simnett, G.M., Howard, R.A., Moses, J.D., Newmark, J.S., Socker, D.G., Halain, J.-P., Defise, J.-M., Mazy, E., Rochus, P.: 2009, The Heliospheric Imagers Onboard the STEREO Mission. *Solar Phys.* **254**, 387. DOI. ADS.

Farrugia, C.J., Jordanova, V.K., Thomsen, M.F., Lu, G., Cowley, S.W.H., Ogilvie, K.W.: 2006, A two-ejecta event associated with a two-step geomagnetic storm. *J. Geophys. Res.* **111**, 11104. DOI. ADS.

Harrison, R.A., Davies, J.A., Möstl, C., Liu, Y., Temmer, M., Bisi, M.M., Eastwood, J.P., de Koning, C.A., Nitta, N., Rollett, T., Farrugia, C.J., Forsyth, R.J., Jackson, B.V., Jensen, E.A., Kilpua, E.K.J., Odstrcil, D., Webb, D.F.: 2012, An Analysis of the Origin and Propagation of the Multiple Coronal Mass Ejections of 2010 August 1. *Astrophys. J.* **750**, 45. DOI. ADS.

Howard, R.A., Moses, J.D., Vourlidas, A., Newmark, J.S., Socker, D.G., Plunkett, S.P., Korendyke, C.M., Cook, J.W., Hurley, A., Davila, J.M., Thompson, W.T., St Cyr, O.C., Mentzell, E., Mehalick, K., Lemen, J.R., Wuelser, J.P., Duncan, D.W., Tarbell, T.D., Wolfson, C.J., Moore, A., Harrison, R.A., Waltham, N.R., Lang, J., Davis, C.J., Eyles, C.J., Mapson-Menard, H., Simnett, G.M., Halain, J.P., Defise, J.M., Mazy, E., Rochus, P., Mercier, R., Ravet, M.F., Delmotte, F., Auchere, F., Delaboudiniere, J.P., Bothmer, V., Deutsch, W., Wang, D., Rich, N., Cooper, S., Stephens, V., Maahs, G., Baugh, R., McMullin, D., Carter, T.: 2008, Sun Earth Connection Coronal and Heliospheric Investigation (SECCHI). *Space Sci. Rev.* **136**, 67. DOI. ADS.

Howard, T.A., Tappin, S.J.: 2009, Interplanetary Coronal Mass Ejections Observed in the Heliosphere: 1. Review of Theory. *Space Sci. Rev.* **147**, 31. DOI. ADS.

Iyemori, T.: 1990, Storm-time magnetospheric currents inferred from mid-latitude geomagnetic field variations. *J. Geomagn. Geoelectr.* **42**, 1249. ADS.

Kaiser, M.L., Kucera, T.A., Davila, J.M., St. Cyr, O.C., Guhathakurta, M., Christian, E.: 2008, The STEREO mission: An introduction. *Space Sci. Rev.* **136**, 5 – 16. doi:DOI.

Kruparova, O., Maksimovic, M., Å AfráNková, J., NěMečEk, Z., Santolik, O., Krupar, V.: 2013, Automated interplanetary shock detection and its application to Wind observations. *J. Geophysical Res.* **118**, 4793. DOI. ADS.

Lepping, R.P., Acũna, M.H., Burlaga, L.F., Farrell, W.M., Slavin, J.A., Schatten, K.H., Mariani, F., Ness, N.F., Neubauer, F.M., Whang, Y.C., Byrnes, J.B., Kennon, R.S., Panetta, P.V., Scheifele, J., Worley, E.M.: 1995, The Wind Magnetic Field Investigation. *Space Sci. Rev.* **71**, 207. DOI. ADS.

Liu, Y.D., Luhmann, J.G., Möstl, C., Martinez-Oliveros, J.C., Bale, S.D., Lin, R.P., Harrison, R.A., Temmer, M., Webb, D.F., Odstrcil, D.: 2012, Interactions between Coronal Mass

Ejections Viewed in Coordinated Imaging and in situ Observations. *Astrophys. J. Lett.* **746**, L15. DOI. ADS.
Liu, Y.D., Luhmann, J.G., Lugaz, N., Möstl, C., Davies, J.A., Bale, S.D., Lin, R.P.: 2013, On Sun-to-Earth Propagation of Coronal Mass Ejections. *Astrophys. J.* **769**, 45. DOI. ADS.
Liu, Y.D., Yang, Z., Wang, R., Luhmann, J.G., Richardson, J.D., Lugaz, N.: 2014, Sun-to-Earth Characteristics of Two Coronal Mass Ejections Interacting Near 1 AU: Formation of a Complex Ejecta and Generation of a Two-step Geomagnetic Storm. *Astrophys. J. Lett.* **793**, L41. DOI. ADS.
Lugaz, N.: 2010, Accuracy and Limitations of Fitting and Stereoscopic Methods to Determine the Direction of Coronal Mass Ejections from Heliospheric Imagers Observations. *Solar Phys.* **267**, 411. DOI. ADS.
Lugaz, N., Farrugia, C.J.: 2014, A new class of complex ejecta resulting from the interaction of two CMEs and its expected geoeffectiveness. *Geophys. Res. Lett.* **41**, 769. DOI. ADS.
Lugaz, N., Farrugia, C.J., Davies, J.A., Möstl, C., Davis, C.J., Roussev, I.I., Temmer, M.: 2012, The Deflection of the Two Interacting Coronal Mass Ejections of 2010 May 23-24 as Revealed by Combined in Situ Measurements and Heliospheric Imaging. *Astrophys. J.* **759**, 68. DOI. ADS.
Maeda, H., Sakurai, K., Ondoh, T., Yamamoto, M.: 1962, A study of solar-terrestrial relationships during the IGY and IGC. *Ann. Geophys.* **18**, 305. ADS.
Minnaert, M.: 1930, On the continuous spectrum of the corona and its polarisation. With 3 figures. (Received July 30, 1930). *Z. Astrophys.* **1**, 209. ADS.
Mishra, W., Srivastava, N.: 2014, Morphological and Kinematic Evolution of Three Interacting Coronal Mass Ejections of 2011 February 13-15. *Astrophys. J.* **794**, 64. DOI. ADS.
Mishra, W., Srivastava, N., Chakrabarty, D.: 2015, Evolution and Consequences of Interacting CMEs of 9 - 10 November 2012 Using STEREO/SECCHI and In Situ Observations. *Solar Phys.* **290**, 527. DOI. ADS.
Mishra, W., Srivastava, N., Davies, J.A.: 2014, A Comparison of Reconstruction Methods for the Estimation of Coronal Mass Ejections Kinematics Based on SECCHI/HI Observations. *Astrophys. J.* **784**, 135. DOI. ADS.
Mishra, W., Srivastava, N., Singh, T.: 2015, Kinematics of interacting CMEs of 25 and 28 September 2012. *J. Geophys. Res.* **120**, 10. DOI. ADS.
Mishra, W., Wang, Y., Srivastava, N.: 2016, On Understanding the Nature of Collisions of Coronal Mass Ejections Observed by STEREO. *Astrophys. J.* **831**, 99. DOI. ADS.
Möstl, C., Farrugia, C.J., Kilpua, E.K.J., Jian, L.K., Liu, Y., Eastwood, J.P., Harrison, R.A., Webb, D.F., Temmer, M., Odstrcil, D., Davies, J.A., Rollett, T., Luhmann, J.G., Nitta, N., Mulligan, T., Jensen, E.A., Forsyth, R., Lavraud, B., de Koning, C.A., Veronig, A.M., Galvin, A.B., Zhang, T.L., Anderson, B.J.: 2012, Multi-point Shock and Flux Rope Analysis of Multiple Interplanetary Coronal Mass Ejections around 2010 August 1 in the Inner Heliosphere. *Astrophys. J.* **758**, 10. DOI. ADS.
Möstl, C., Rollett, T., Frahm, R.A., Liu, Y.D., Long, D.M., Colaninno, R.C., Reiss, M.A., Temmer, M., Farrugia, C.J., Posner, A., Dumbović, M., Janvier, M., Démoulin, P., Boakes, P., Devos, A., Kraaikamp, E., Mays, M.L., Vršnak, B.: 2015, Strong coronal channelling and interplanetary evolution of a solar storm up to Earth and Mars. *Nature Comm.* **6**, 7135. DOI. ADS.
Munro, R.H., Gosling, J.T., Hildner, E., MacQueen, R.M., Poland, A.I., Ross, C.L.: 1979, The association of coronal mass ejection transients with other forms of solar activity. *Solar Phys.* **61**, 201. DOI. ADS.
Ogilvie, K.W., Chornay, D.J., Fritzenreiter, R.J., Hunsaker, F., Keller, J., Lobell, J., Miller, G., Scudder, J.D., Sittler, E.C. Jr., Torbert, R.B., Bodet, D., Needell, G., Lazarus, A.J., Steinberg, J.T., Tappan, J.H., Mavretic, A., Gergin, E.: 1995, SWE, A Comprehensive Plasma Instrument for the Wind Spacecraft. *Space Sci. Rev.* **71**, 55. DOI. ADS.
Poland, A.I., Howard, R.A., Koomen, M.J., Michels, D.J., Sheeley, N.R. Jr.: 1981, Coronal transients near sunspot maximum. *Solar Phys.* **69**, 169. DOI. ADS.
Sheeley, N.R. Jr., Herbst, A.D., Palatchi, C.A., Wang, Y.-M., Howard, R.A., Moses, J.D., Vourlidas, A., Newmark, J.S., Socker, D.G., Plunkett, S.P., Korendyke, C.M., Burlaga, L.F., Davila, J.M., Thompson, W.T., St Cyr, O.C., Harrison, R.A., Davis, C.J., Eyles, C.J., Halain, J.P., Wang, D., Rich, N.B., Battams, K., Esfandiari, E., Stenborg, G.: 2008, Heliospheric Images of the Solar Wind at Earth. *Astrophys. J.* **675**, 853. DOI. ADS.
Shen, C., Wang, Y., Wang, S., Liu, Y., Liu, R., Vourlidas, A., Miao, B., Ye, P., Liu, J., Zhou, Z.: 2012, Super-elastic collision of large-scale magnetized plasmoids in the heliosphere. *Nature* **8**, 923. DOI. ADS.

Stone, E.C., Frandsen, A.M., Mewaldt, R.A., Christian, E.R., Margolies, D., Ormes, J.F., Snow, F.: 1998, The Advanced Composition Explorer. *Space Sci. Rev.* **86**, 1. DOI. ADS.

Temmer, M., Vršnak, B., Rollett, T., Bein, B., de Koning, C.A., Liu, Y., Bosman, E., Davies, J.A., Möstl, C., Žic, T., Veronig, A.M., Bothmer, V., Harrison, R., Nitta, N., Bisi, M., Flor, O., Eastwood, J., Odstrcil, D., Forsyth, R.: 2012, Characteristics of Kinematics of a Coronal Mass Ejection during the 2010 August 1 CME-CME Interaction Event. *Astrophys. J.* **749**, 57. DOI. ADS.

Thernisien, A.F.R., Howard, R.A., Vourlidas, A.: 2006, Modeling of Flux Rope Coronal Mass Ejections. *Astrophys. J.* **652**, 763. DOI. ADS.

Vemareddy, P., Mishra, W.: 2015, A Full Study on the Sun-Earth Connection of an Earth-directed CME Magnetic Flux Rope. *Astrophys. J.* **814**, 59. DOI. ADS.

Vourlidas, A., Subramanian, P., Dere, K.P., Howard, R.A.: 2000, Large-Angle Spectrometric Coronagraph Measurements of the Energetics of Coronal Mass Ejections. *Astrophys. J.* **534**, 456. DOI. ADS.

Zurbuchen, T.H., Richardson, I.G.: 2006, In-Situ Solar Wind and Magnetic Field Signatures of Interplanetary Coronal Mass Ejections. *Space Sci. Rev.* **123**, 31. DOI. ADS.